# Development of a Cesium Fountain Clock at HUST: Preliminary Results

Hui Li, Yuanbo Du, Hongli Liu, Yunyi Guo, Mingming Liu, Wenbing Li, Zehuang Lu

***Abstract*—A cesium atomic fountain clock is under development at Huazhong University of Science and Technology (HUST) in China. In this paper, we describe the construction of the entire fountain clock system and report the preliminary results. A frequency stability of $2.5\times10^{-13}\tau^{-1/2}$ has been achieved by inter-comparison with a hydrogen maser, and the factors limiting the frequency stability are also discussed.**

## I. Introduction

CESIUM fountain clocks serve as primary frequency standards for International Atomic Time (TAI) [1-12], and have many applications in fundamental research fields such as precision measurement physics [13-14], relativity theory testing [15], as well as applications in navigation systems [16]. Many laboratories around the world, including LNE-SYRTE, NIST, PTB, NPL, NIM, have developed their own fountain clocks, with excellent results [17]. The frequency stability and systematic uncertainty (type B uncertainty) are two key indicators of the performance of a fountain clock. The best frequency stability of $1.6\times10^{-13}\tau^{-1/2}$ is measured at LNE-SYRTE [8], and a systematic uncertainty of $1.1\times10^{-16}$ has been reported by NIST [2]. A cesium fountain clock is currently under development at HUST in China to provide a local time-frequency reference for the National Precise Gravity Measurement Facility (PGMF) that is under construction at HUST [18]. Meanwhile, it is also used to measure the absolute clock transition frequency of $^{27}Al^+$ ion optical clock that is under development in our group [19].

During the construction of the cesium fountain clock, we have completed the vacuum physics package, the optical system, the microwave synthesizer, and the electronics control system. A detailed description of the fountain clock system and the first evaluation result of the frequency stability is presented in this paper. We will first start with the description of the cesium fountain clock system, including the vacuum physics package, the optical system, the microwave synthesizer, and the electronic control system, followed with experimental results, including measurement of time of flight (TOF) signals, recording of Ramsey fringes, and servo locking of the clock transition. Finally we conclude with a conclusion.

## II. Description of the Cesium Fountain Clock System

### *A. Vacuum physics package*

Figure 1(a) shows the schematic drawing of the cesium fountain clock physics package. The whole structure is about 2.2 meters high, and is supported by four stages of aluminum plates. The main structure of the vacuum physics package is made of non-magnetic titanium alloy material. In order to facilitate the transportation of the cesium fountain clock between different labs, four casters are installed at the bottom of the package. At the bottom of the package, an additional three-point adjustable support structure is added. The three-point structure is used to support and fix the entire physics package, and is also used to adjust the levelling of the vacuum physics package before fixing the entire package to the ground. The vacuum physics package is composed of magnetic shielding, C-field coil, upper vacuum chamber, Ramsey cavity, state selection cavity, state detection zone, 3D-MOT zone, 2D-MOT zone, and equipped with a 75 l/s ion pump and a 20 l/s ion pump to ensure a high vacuum environment inside the vacuum physics package.

An angle-adjustment unit located in the 3D-MOT zone is used to tune the atomic cloud launching directions by adjusting the angles of the 3D-MOT cavity. The design of the angle-adjustment unit is similar to that of PTB-CSF2 [20]. The launching directions of cesium atoms can be independently adjusted to increase the number of the falling-back atoms. In order to avoid the influence of environmental magnetic field on fountain operations, a four-layer magnetic shield made of mu metal is placed outside the vacuum chamber and the shielding factor is about $10^4$. A homogeneous C-field of about 127 nT is produced by a C-Field solenoid with a total length of 1110 mm. The Ramsey cavity with four-port microwave feed is integrated with the state selection cavity with one-port microwave feed, and the $Q_{loaded}$ factor of the Ramsey cavity is approximately

This work is supported by the National Natural Science Foundation of China (Grant Nos. 11804108, 11774108, and 11904113), the National Key Research and Development Program of China (Grant No. 2017YFA0304400), and the China Postdoctoral Science Foundation (Grant No. 2017M612431). (Corresponding authors: Zehuang Lu and Yuanbo Du)

Hui Li, Hongli Liu, Yunyi Guo and Mingming Liu are with the School of Physics, Huazhong University of Science and Technology, China. (e-mail: hui_li@hust.edu.cn)

Yuanbo Du is currently an associate professor at the School of Physics and Astronomy, Sun Yat-sen University, China. (e-mail: duyb6@mail.sysu.edu.cn)

Wenbing Li is doing postdoctoral research at Helmholtz-Institut Mainz, Mainz, Germany.

Zehuang Lu was a research group leader at Max-Planck Institute for the Science of Light, Germany from 2004 to 2010. He is now a professor at Huazhong University of Science and Technology. (e-mail: zehuanglu@hust.edu.cn)

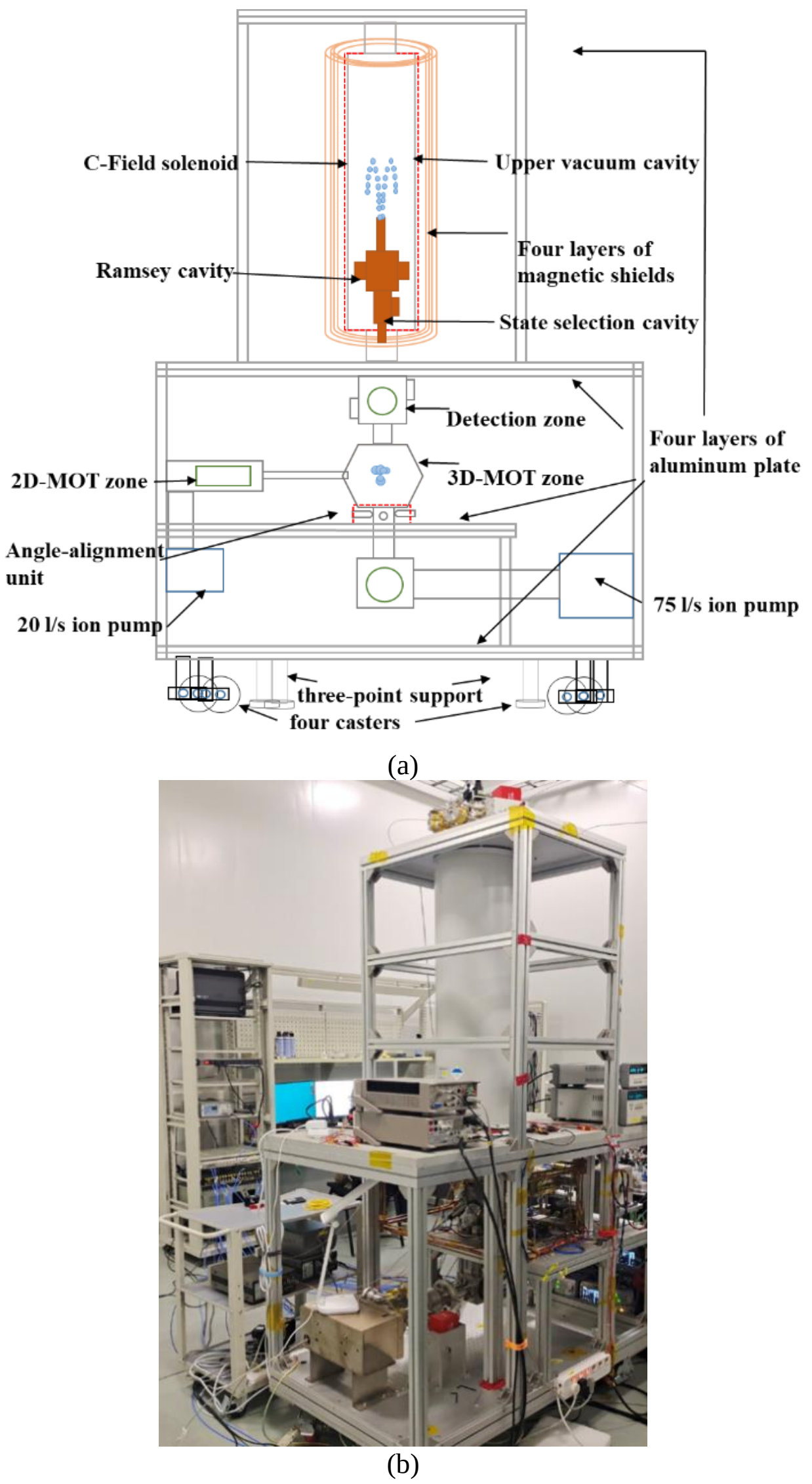

Fig. 1. (a) The schematic drawing of the cesium fountain clock physics package. (b) A picture of the cesium fountain clock under development at HUST.

13500. The four-port microwave feed used on the Ramsey cavity can reduce the distributed cavity phase (DCP) shift [21] of the cesium fountain clock. The TOF signals of the atoms are detected in the state detection zone.

A 2D-MOT [22] is designed to increase the efficiency of loading cesium atoms, but it is not used in the experiments at the moment, and the trapping and cooling of cesium atoms is completed in the 3D-MOT zone. About $10^8$ cesium atoms are captured in the 3D-MOT zone with six cooling laser beams, one repump laser beam. The magnetic field gradient is 5.6 G/cm. The cesium atoms are launched with 4.43 m/s in moving optical molasses and the atoms are cooled to around 4.3 μK by polarization gradient cooling in the final stage.

## *B. Optical system*

The optical system shown in Fig. 2 can be divided into two parts. One is used to provide the cooling light, the push light, and the probe light, and the other one is used to provide the repump light. Two Toptica DL Pro 850 semiconductor lasers and a Toptica BOOSTA Pro 850 tapered amplifier are used in the optical path. According to the optical power demand of the system, a semiconductor laser and a tapered amplifier is used to provide the cooling light, the push light and the probe light, and the output power is about 1.5 W after being amplified by the tapered amplifier. The repump laser is provided by another semiconductor laser. The powers and frequencies of different light beams from the optical system are controlled by acousto-optic modulators (AOMs). The repump laser is stabilized to the $^{133}$Cs $6^2S_{1/2}|F{=}3\rangle \rightarrow 6^2P_{3/2}|F{=}4\rangle$ transition using saturation absorption technique. Other laser beams, including the cooling light, the push light, and the probe light, are locked to the repump laser by using a frequency voltage conversion circuit and a servo locking loop, and the laser frequency is resonant with the $^{133}$Cs $6^2S_{1/2}|F{=}4\rangle \rightarrow 6^2P_{3/2}|F{=}5\rangle$ transition.

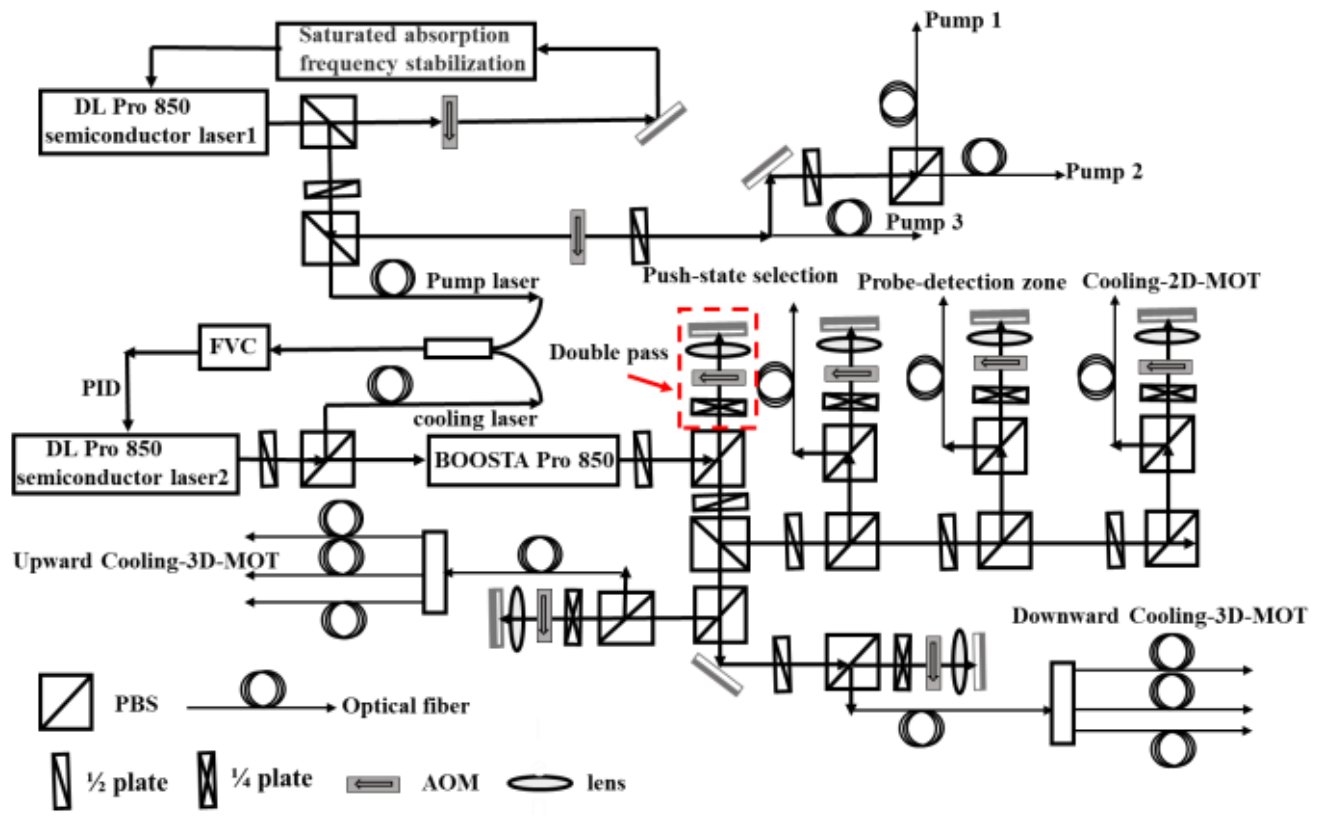

Fig. 2. Schematic of the optical system.

The cooling light ($\sigma^+$-$\sigma^-$) and the repump light with a diameter of about 20 mm are delivered to the 3D-MOT zone by fiber beam expanders. The probe beam is divided into two parallel beams in the state detection zone and reflected by a mirror to form standing light wave. The two probe beams are used to detect the $6^2S_{1/2}|F{=}4, m_F{=}0\rangle$ atoms and the $6^2S^{1/2}|F{=}3, m_F{=}0\rangle$ atoms, respectively. One of the probe beams is partially blocked as the push light to blow away the $6^2S_{1/2}|F{=}4, m_F{=}0\rangle$ atoms after the first detection. The other probe beam combined with the repump light is used to detect the $6^2S^{1/2}|F{=}3, m_F{=}0\rangle$ atoms. One more probe beam is used to push away the $6^2S_{1/2}|F{=}4, m_F{\neq}0\rangle$ atoms in the state selection experiment.

## *C. Microwave synthesizer*

Two microwave synthesizers based on 5 MHz Oven Controlled Crystal Oscillator (OCXO), 100 MHz Voltage Controlled Crystal Oscillator (VCXO) and Direct Digital Frequency Synthesis (DDS) are built in our lab. The microwave frequency signal of 9.192631770 GHz generated by the microwave synthesizer is used as the local oscillator for the HUST fountain clock.

A detailed schematic diagram of the microwave synthesizer is shown in Fig. 3. In the first phase locked loop (PLL1), the frequency of 5 MHz OCXO (Microsemi 1000C) is frequency-multiplied to 100 MHz, then the 100 MHz is divided into two branches through the power divider. The first branch is locked to a Hydrogen maser (iMaser 3000) by PLL1, the other one is phase locked with the 100 MHz signal generated by the voltage control oscillator (Voltage Control Crystal Oscillator, VCXO) in PLL2. The 100 MHz signal from PLL2 is frequency-doubled to 200 MHz, and then divided into three branches by a power splitter. One of them is used as the reference for a Direct Digital Frequency Synthesis (DDS) with a frequency at 40 MHz after a divided-by-five frequency divider. The other one is used to drive a non-linear transmission line (NLTL) comb generator to generate harmonic signal, and the 9.0 GHz is obtained after a band-pass filter (BPF). The frequency of 7.368 MHz is generated by mixing the 9.0 GHz signal with the 8.9926 GHz signal generated by a dielectric oscillator (DRO). Then this 7.368 MHz signal and the 7.368 MHz signal generated by the DDS are phase detected by a phase-frequency detector (PFD). The frequency of the 8.9926 GHz generated by the DRO is locked in PLL3. The last one of the 200 MHz signal after going through an interferometric switch is mixed with the 8.9926 GHz signal from PLL3 to produce the 9.1926 GHz clock transition

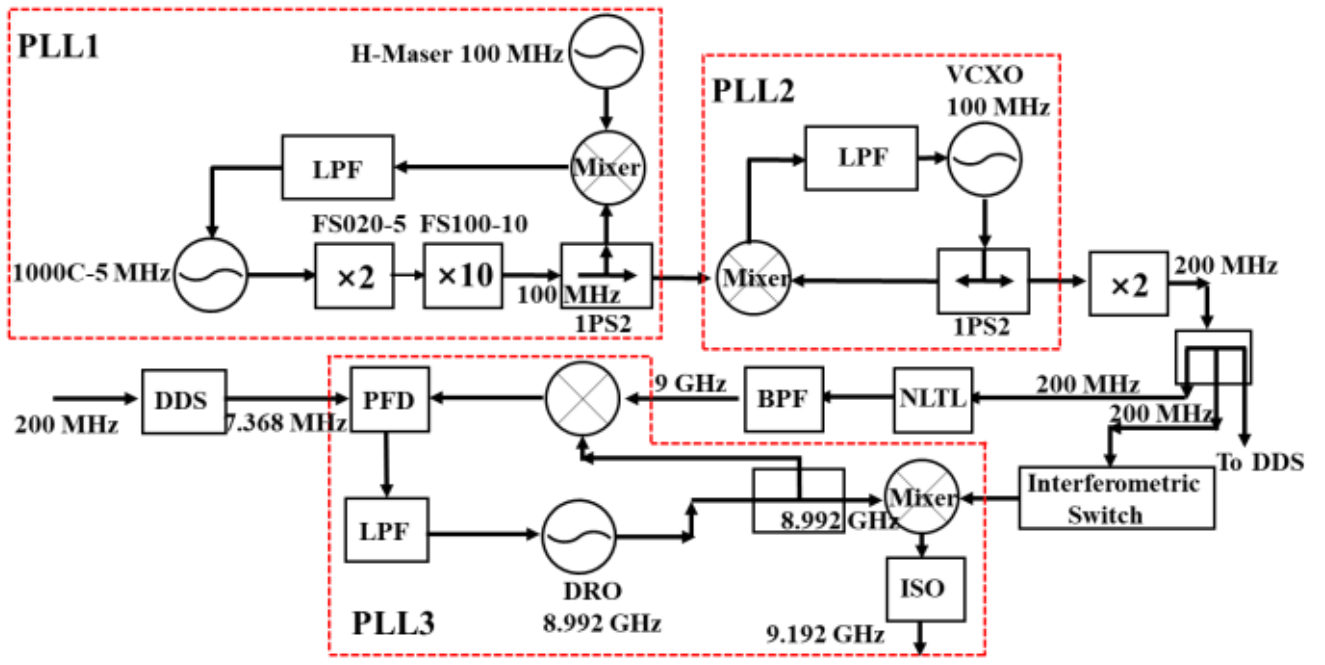

Fig. 3. Schematic diagram of the microwave synthesizer.

signal.

Figure 4 shows the results of the absolute phase noise and the residual phase noise of the microwave synthesizer measured in the experiment. The short frequency stability of the cesium fountain clock is limited by the quantum projection noise [23] and the local oscillator noise (Dick effect) [24]. The local oscillator noise comes from the phase noise of the microwave source, and the contribution of the local oscillator noise introduced by Dick effect to the frequency stability of the cesium fountain clock is calculated to be $1.53\times10^{-13}\tau^{-1/2}$. Constructing a microwave synthesizer with lower phase noise and increasing the number of detected atoms can improve the frequency stability of fountain clocks. The influence of the microwave synthesizer based on photonic microwave generation [25] or cryogenic sapphire oscillator (CSO) [26] on the frequency stability of fountain clocks is at $10^{-16}\tau^{-1/2}$ level, which can be ignored in comparison with the quantum projection noise. Our laboratory has also completed the construction of a photonic microwave generator, which will be put into use soon.

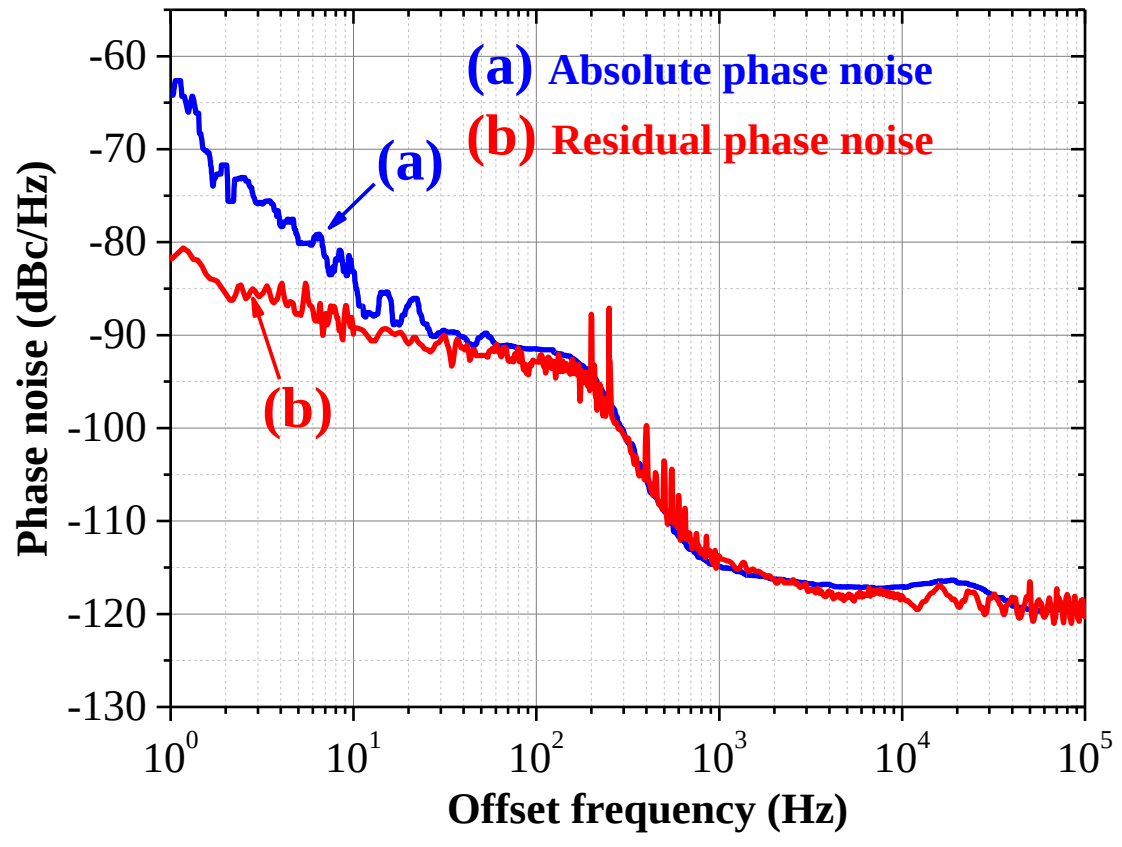

Fig. 4. Absolute phase noise and residual phase noise of the microwave synthesizer.

### *D. Electronic control system*

The program of the electronic control system of our cesium fountain clock is written in LabVIEW, which has two main functions of control and acquisition. The hardware includes analog output board and data acquisition board. The analog output board (NI PCI-6733) provides 16 channels of analog voltage signal (trigger signal) to control the microwave switches, optical shutters, magnetic field switch, and temperature control switch, so that different stages of experiments can be successfully completed. The data acquisition board perform the function of data monitoring and collection. It can monitor the power of laser beams, the temperature of the vacuum chamber, and the long-term changes of the magnetic field inside the physical package. Also, it is used to record the TOF signals and the Ramsey fringes

Figure 5 shows the control sequence of the cesium fountain clock. In the 3D-MOT stage, the magnetic field $B_{MOT}$, cooling light and repump light are ON at the same time to trap cesium atoms, and the repump light pumps the atoms from $6^2S_{1/2}|F=3>$ to $6^2P_{3/2}|F=4>$ to ensure continuous cooling of the atoms. The whole process lasts for 1 s. Then the magnetic field is turn off, and the atoms enter a 60 ms of optical molasses process (including moving optical molasses). By controlling AOM to change the relative frequency detuning (6 MHz) of upward cooling beams and downward cooling beams, the atoms can obtain an initial launching speed of 4.43 m/s. The atoms in $6^2S_{1/2}|F=3>$ state Zeeman sublevels after optical molasses are launched into the state selection cavity for atomic state selection, and the atoms in $6^2S_{1/2}|F=3, m_F=0>$ state are selected by a microwave π-pulse on atoms, and the remaining atoms are blow away by the push laser for 2 ms. After the state selection, the atoms enter the Ramsey cavity and interacts with a microwave $\pi$/2-pulse twice, each time for 60 ms, and the free evolution time between the two interactions is 540 ms. The duration of two-level detection is about 40 ms, and the fountain cycle is 2.5 s.

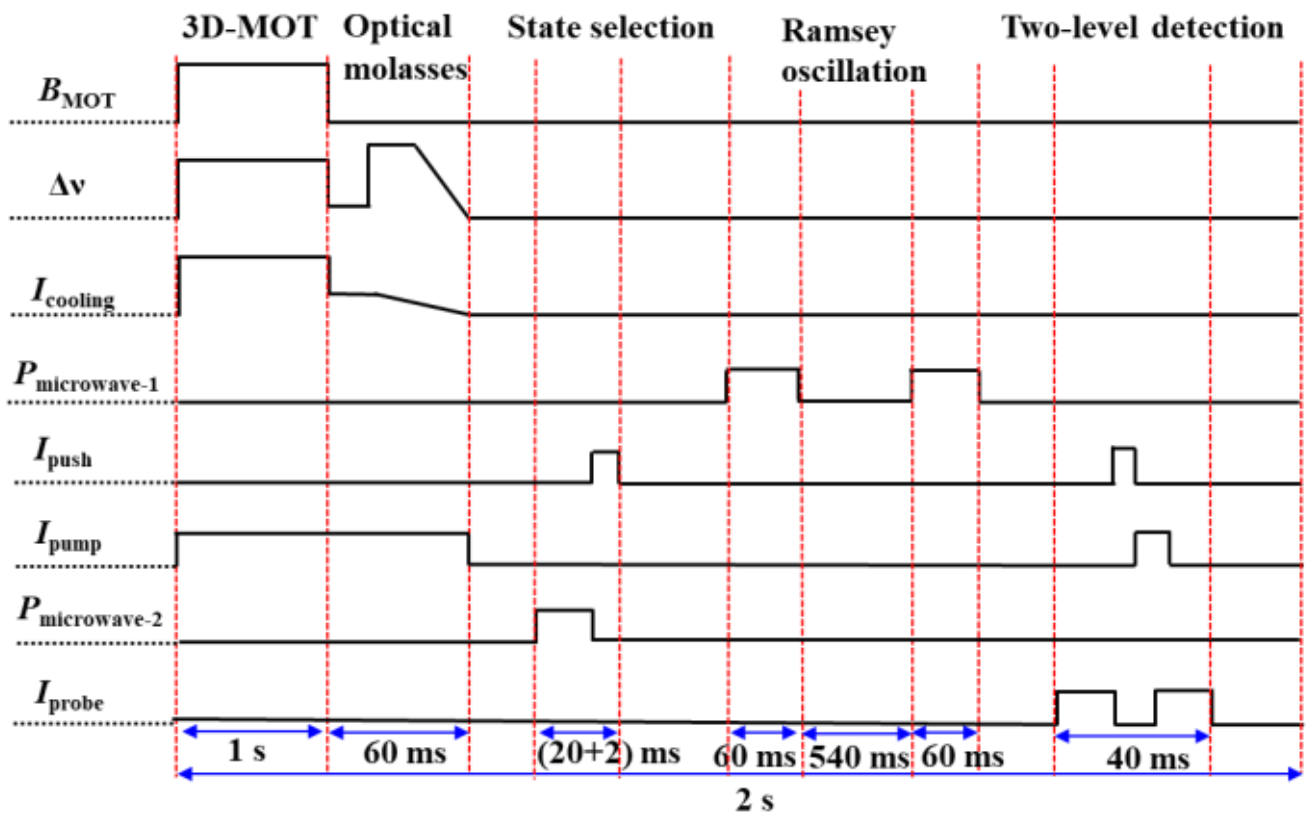

Fig. 5. Control sequence of the electronic control system in the cesium fountain clock.

## III. Experimental Results

### A. *The C field mapping*

Atoms are easily disturbed by the environmental magnetic field when interacting with microwaves in the microwave cavity, and the environmental magnetic field mainly comes from the Earth magnetic field. Second-order Zeeman shift due to the Earth magnetic field is a major contribution to the frequency uncertainty of cesium fountain clocks.

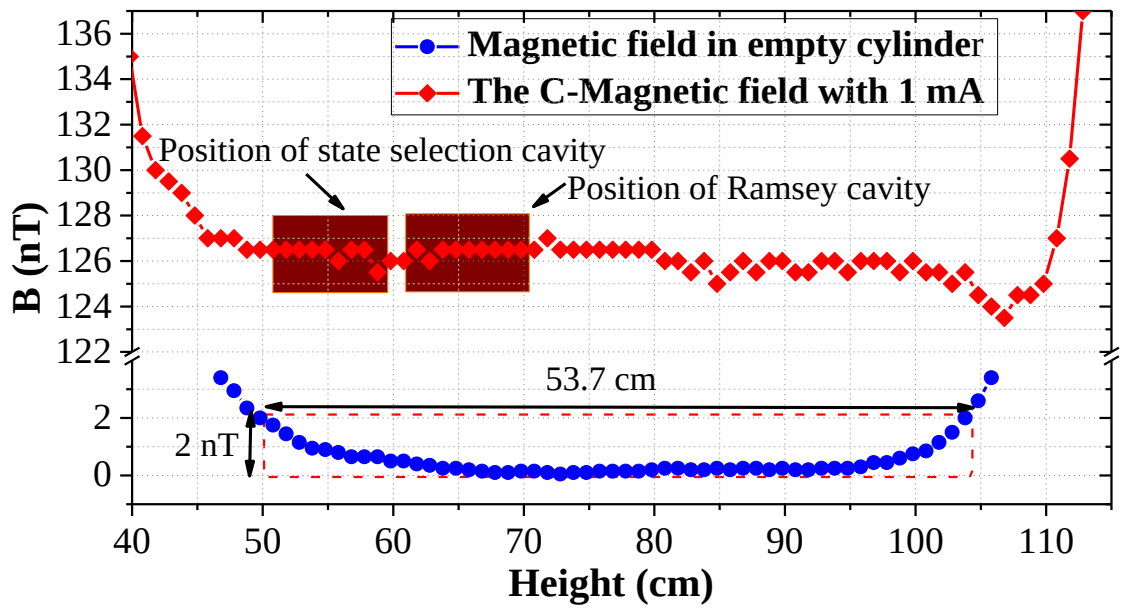

Fig. 6. Magnetic field with and without C field coil current along the axis of the vacuum physics package.

To ensure the atoms are not disturbed by the environmental magnetic field in the process of interacting with microwave signal in the Ramsey cavity, a four-layer cylindrical magnetic shield is installed, which can shield the external environmental magnetic field to nT level. Fig. 6 shows the shielding results of the magnetic shielding cylinder. We measured the magnetic field in the shielding enclosure before sealing the vacuum. The homogeneity length of the magnetic field is 53.7 cm, and the fluctuation of the magnetic field is less than 2 nT. As we can see from Fig. 6, when the C field coil is loaded with 1 mA current, the average magnetic field at the positions where the state selection cavity and the Ramsey cavity are located are both about 126.5 nT, and the magnetic field fluctuation does not exceed 2 nT. The measured results showed that a good magnetic field environment is provided for the experiment, and the current we chose for the experiment was set at 1mA.

### B. *TOF signal of atomic cloud*

About $10^8$ atoms are trapped in the 3D-MOT, and nearly 8% of the atoms are detected after launching and falling back. Considering that the size of the atomic cloud becomes larger due to the thermal diffusion in the process of launching up and falling down, the size of the detection beam in the state detection zone is larger than the size of the atomic cloud. The TOF signals of the falling back atoms are detected by a square probe beam (15 mm×4 mm) in the state detection zone. Two square probe beams are arranged in the detection zone. The first probe light is above the second probe light and there are a push light (15 mm×2 mm) and a repump light (15 mm×2 mm) between the two probe lights. The push light and the repump light are used in two-level detection after Ramsey oscillation.

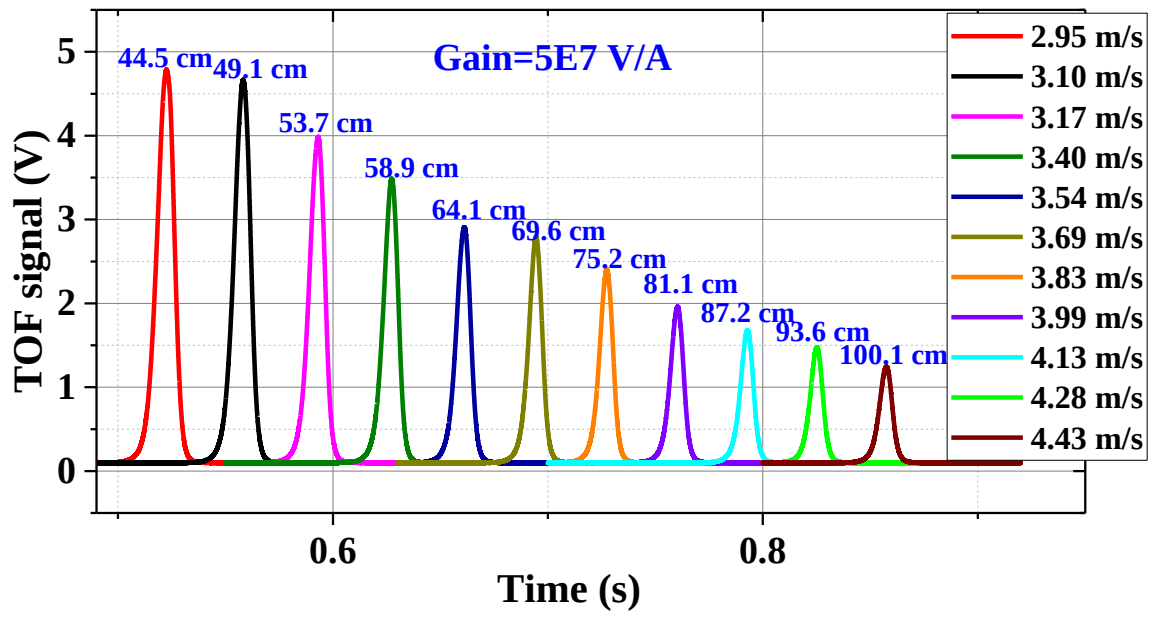

Fig. 7. The TOF signals of the falling back atoms with different initial launching speeds.

Figure 7 shows the TOF signals of the falling back atoms detected by the first probe light with different initial launching speeds. The higher the launching height of the atomic cloud, the less the number of the falling back atoms. When the atomic cloud is tossed to a height of 49.1 cm, it began to enter the Ramsey cavity, and it flew out of the cut-off waveguide at the top of the Ramsey cavity at a height of 76.4 cm. We chose a height of 100.1 cm for the operation of the fountain clock.

### C. *State Preparation and Microwave spectra of Zeeman levels*

The power of the 9.192631770 GHz microwave signal is changed by controlling the digital attenuator of the microwave synthesizer. The state selection cavity Rabi oscillation curve is shown in Fig. 8(a), and the power scanning range is from -80 dBm to -40 dBm. At the first peak, the atom has the first full transition of $6^2S_{1/2}|F=4, m_F=0\rangle \rightarrow 6^2S_{1/2}|F=3, m_F=0\rangle$, and the microwave pulse area is $\pi$. The microwave pulse areas corresponding to the following peaks in Fig. 8(a) are 3$\pi$, 5$\pi$, 7$\pi$, etc. We choose the microwave power value corresponding to the first peak for state selection. Fig. 8(b) shows the microwave spectra of Rabi oscillation varies with microwave frequency in the state selection cavity. Seven Zeeman level transition lines of $\Delta F=\pm1$, $\Delta m_F=0$ are observed in Fig. 8(b), and weaker transition lines of $\Delta F=\pm1$, $\Delta m_F=\pm1$ also exist, indicating there are still small magnetic field components that are perpendicular to the direction of the C field. The transition line strengths of $\Delta F=\pm1$, $\Delta m_F=\pm1$ lines are negligible compared with that of the $6^2S_{1/2}|F=4, m_F=0\rangle \rightarrow 6^2S_{1/2}|F=3, m_F=0\rangle$ transition.

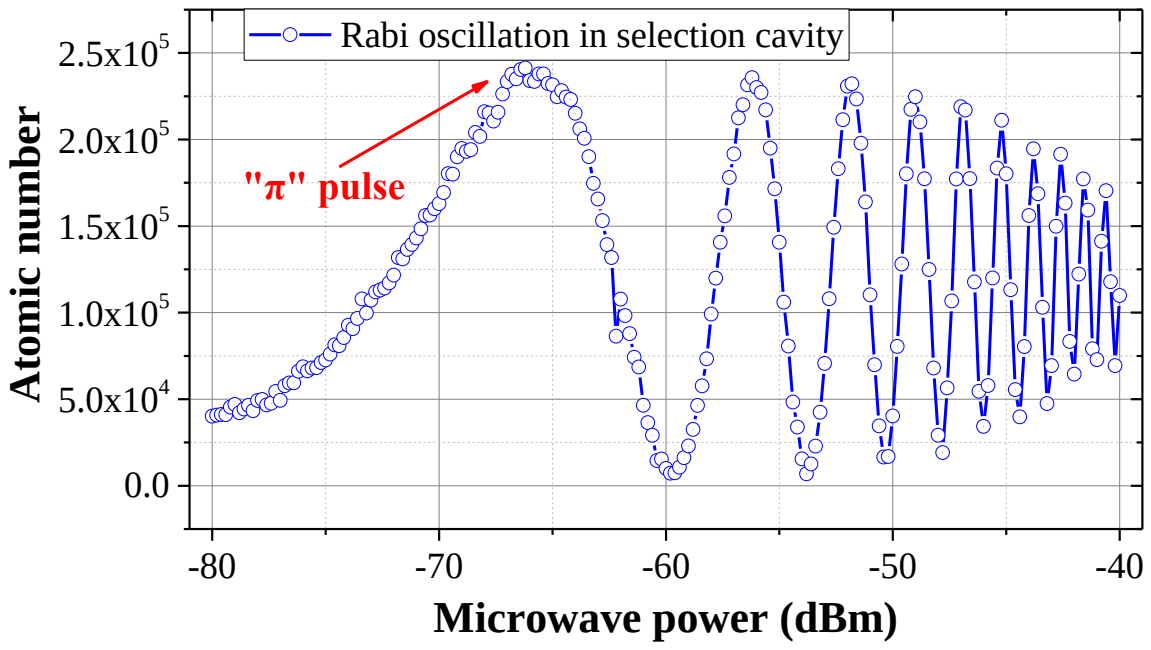

(a)

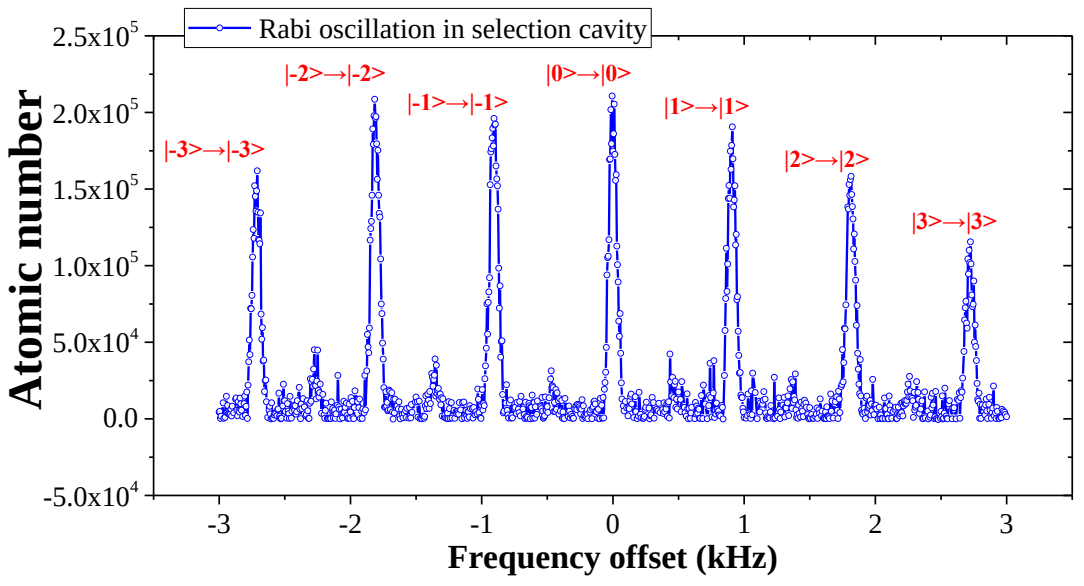

(b)

Fig. 8. (a) Rabi oscillation curve varies with microwave power during state selection in the state selection cavity. (b) Microwave spectra of Rabi oscillation varies with microwave frequency in the state selection cavity with state selection.

### *D. Observation of Ramsey fringes*

The cesium atomic cloud continues travels up after the state selection, goes through the Ramsey cavity, and then falls back through the Ramsey cavity again. The atomic cloud passes through the Ramsey cavity twice, and there is a free evolution time between flying out of the Ramsey cavity and falling back to the Ramsey cavity. The microwave frequency feed into the Ramsey cavity is fixed at 9.192631770 GHz, and the Ramsey oscillation as a function of microwave power is obtained in Fig. 9. The microwave power is scanned from -80 dBm to -30 dBm, and the step is 0.2 dB. The microwave power corresponding to the first peak in Figure 9 indicates that the interaction area between atomic cloud and microwave, and the power of the first peak for the chosen "π/2" area is -62.9 dBm.

On the way down, the TOF signal of the $6^2S_{1/2}|F=4, m_F=0>$ atoms is detected by the first probe light in the detection zone, then the remaining atoms are blown away by the push light. The $6^2S_{1/2}|F=3, m_F=0>$ atoms are pumped to the $6^2P_{3/2}|F=4, m_F=0>$ by the repump light and then detected by the second probe light. The number of atoms in the $6^2S_{1/2}|F=4, m_F=0>$ is $N_4$, and the number of atoms in $6^2S_{1/2}$ $|F=3, m_F=0>$ is $N_3$. The ratio $N_4/(N_3+N_4)$ is the transition probability of the cesium fountain clock transition. Ramsey fringes are obtained by varying the microwave frequency and recording the corresponding transition probabilities.

Figure 10 shows the Ramsey fringes measured in our lab. The fringe contrast is over 90%, and the full width at half maximum (FWHM) of the central Ramsey fringe is about 0.92 Hz, as shown in the inset. The height of the launch is around one meter from the center of the 3D-MOT zone. The initial launching speed of atoms is 4.43 m/s. The initial temperature of atoms is around 4 μk.

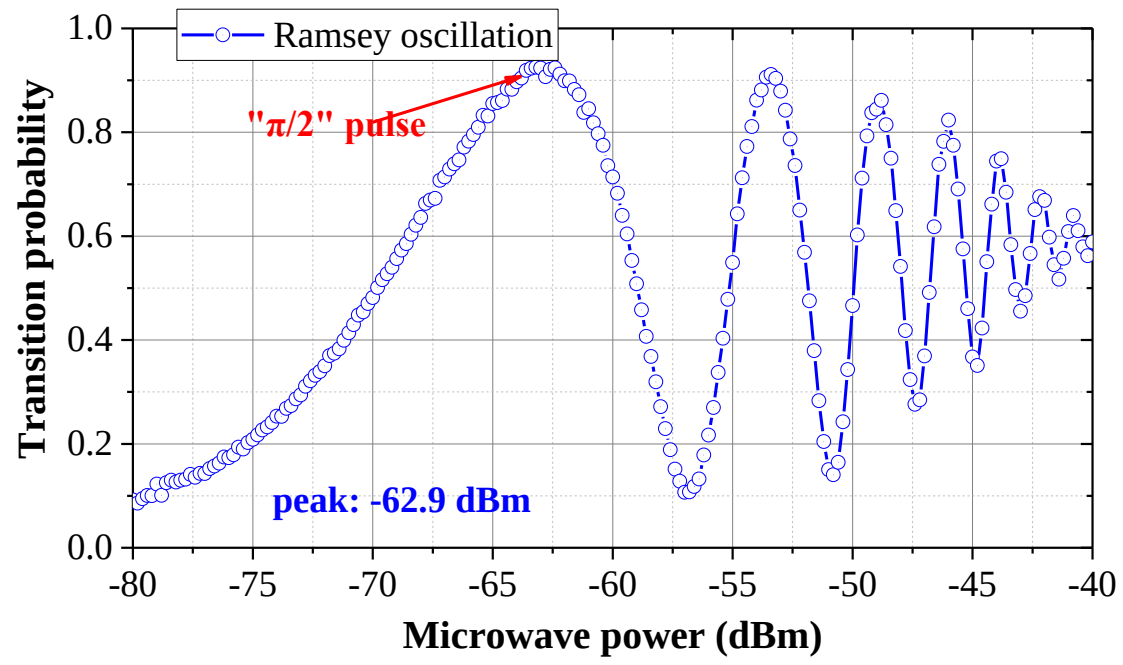

Fig. 9. Variation of Ramsey oscillation transition probability with microwave power in Ramsey cavity.

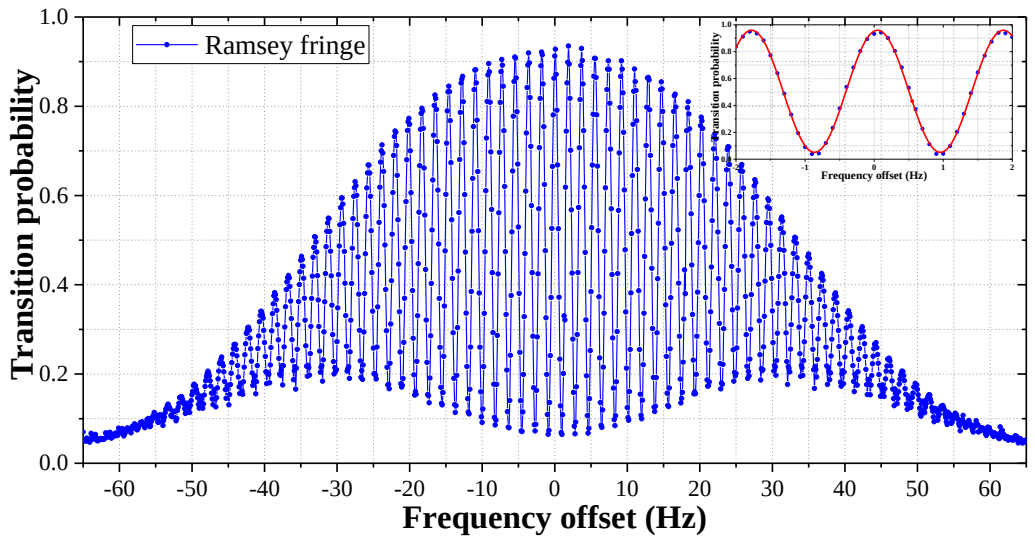

Fig. 10. Ramsey fringes measured in our lab. Inset: the central Ramsey fringes.

### *E. Fountain clock frequency stability*

Cesium fountain clocks need to operate in closed-loop to output stable clock signals. The microwave frequency is locked to the FWHM of the central Ramsey fringe, and the frequency is square-wave modulated at $\nu$-$\Delta\nu/2$ and $\nu$+$\Delta\nu/2$ in each cycle of the cesium fountain operation where $\Delta\nu$ is the FWHM of the central Ramsey fringe and $\nu$ is the central microwave frequency of 9.192631770 GHz.

Figure 11 is typical TOF signals for the falling back atoms in state $6^2S_{1/2}|F=4, m_F=0>$ and $6^2S_{1/2}|F=3, m_F=0>$ obtained during the frequency modulation process. The transition probability $P=N_4/(N_3+N_4)$ of $6^2S_{1/2}|F=4, m_F=0>\rightarrow 6^2S_{1/2}|F=3, m_F=0>$is measured from each cycle of the cesium fountain. The error

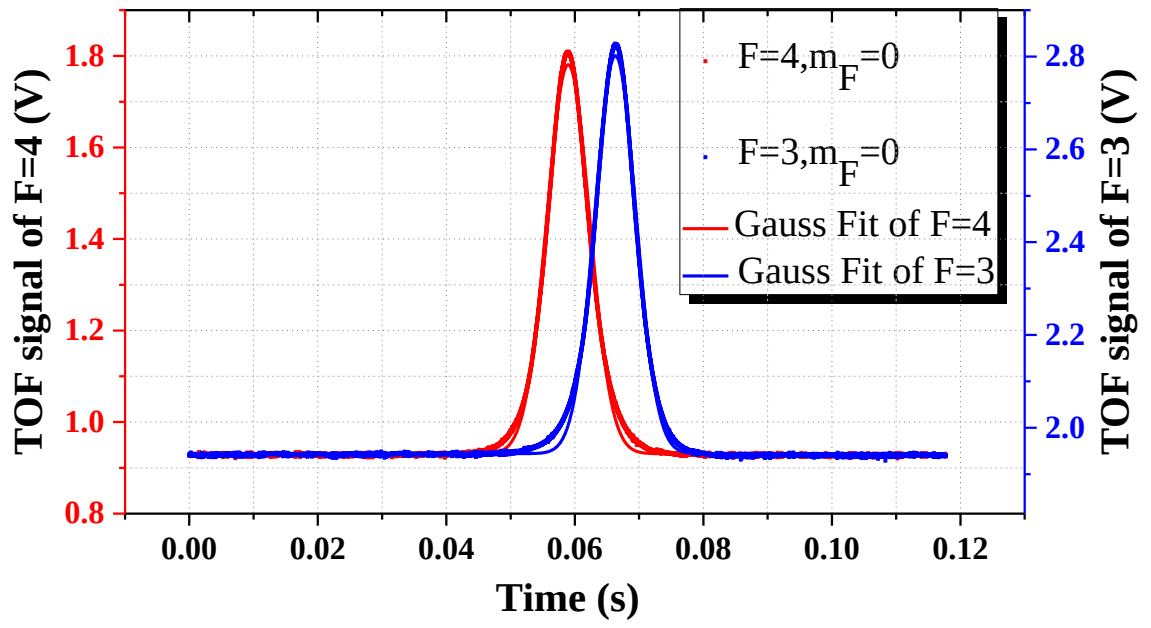

Fig. 11. The TOF signals. The red dot is the TOF signal of $6^2S_{1/2}|F=4, m_F=0>$ and the red line is the fitting curve, the blue point is the TOF signal of $6^2S_{1/2}|F=3, m_F=0>$ and the blue line is the fitting curve.

signal can be obtained from the transition probability $P$ and fed back to the microwave oscillator.

The cesium fountain clock is operated with a cycle time of 2.5 s. Fig. 12 shows the measured Allan deviation of the frequency stability of the HUST fountain clock compared with a hydrogen maser (iMaser 3000). The result of the frequency stability of a hydrogen maser (iMaser 3000) against another hydrogen maser (MHM2010) is also given in Fig. 12. The frequency stability of the HUST fountain clock is measured to be $2.5\times10^{-13}\tau^{-1/2}$.

The main factors affecting the frequency stability of the fountain clock include quantum projection noise, detection noise, photon shot noise and the local oscillator noise (Dick effect). The influence of these noise contributions on the frequency stability can be written as

$$\sigma(\tau)=\frac{1}{\pi Q_{at}}\sqrt{\frac{T_c}{\tau}}\left(\frac{1}{N_{at}}+\frac{1}{N_{at}\varepsilon\eta_{ph}}+\frac{2\sigma_{\delta N}^2}{N_{at}^2}+\gamma\right)^{1/2}. \quad (1)$$

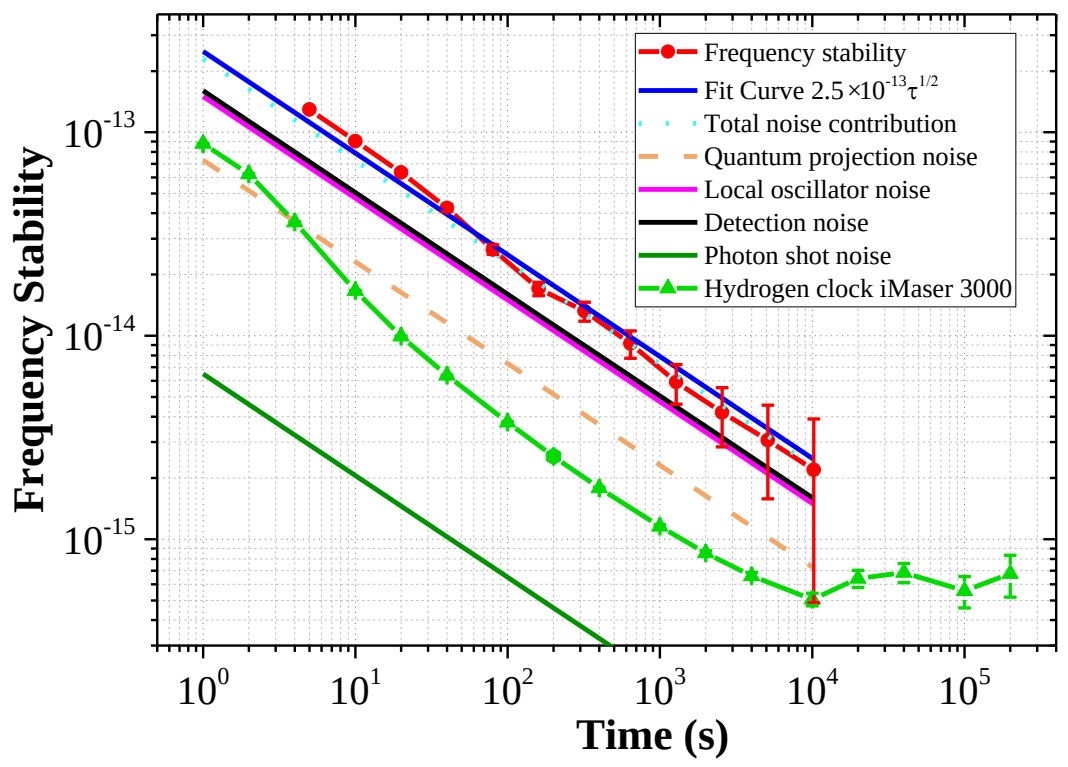

Fig. 12. Frequency stability of the cesium fountain clock at HUST.

In Eq. (1), $Q_{at}$ is the quality factor of the central Ramsey fringe, $\tau$ is the measurement time, $T_c$ is the cycle time of the fountain, $N_{at}$ is the number of detected atoms, $\eta_{ph}$ is the number of photons produced by an atom during the detection process, $\varepsilon$ is the photon collection efficiency, $\sigma_{\delta N}$ is the uncorrelated root mean square fluctuation of the atomic number in each detection channel, $\gamma$ represents the contribution of frequency noise from the local oscillator. The total noise contributions has been summarized in Table 1. The predicted result is consistent with the measured result, and the frequency stability is limited by the detection noise and the local oscillator noise (Dick effect) according to Table 1. The different noise contributions are also plotted in Fig. 12.

TABLE I TOTAL NOISE CONTRIBUTION OF CLOCK STABILITY

| Noise contributions | $\sigma(\tau)$ ($\tau^{-1/2}$) |
|---|---|
| Quantum projection noise | $7.34\times10^{-14}$ |
| Detection noise | $1.55\times10^{-13}$ |
| Photon shot noise | $6.45\times10^{-15}$ |
| Local oscillator noise | $1.53\times10^{-13}$ |
| Total(predicted) | $2.30\times10^{-13}$ |
| Total(measured) | $2.50\times10^{-13}$ |

## IV. Conclusion

We have built a cesium fountain clock at HUST and completed the evaluation of the frequency stability by comparing it with a hydrogen maser. The contrast of the Ramsey fringe is over 90%, and the FWHM of the central Ramsey fringe is 0.911 Hz. The frequency stability of the cesium fountain clock is $2.5\times10^{-13}\tau^{-1/2}$, and the main contributions are also analyzed and discussed.

The short-term frequency stability of the cesium fountain clock can be improved by replacing the microwave source using photonic microwave source. The photonic microwave source is under development in our lab and we plan to replace it in further experiments. Major uncertainty contributions are being evaluated, and we expect more detailed results in the future.

## Acknowledgment

We acknowledge fruitful discussions with professor Minkang Zhou and Rong Wei.